\documentclass[conference]{IEEEtran}

\usepackage{mathrsfs}

\ifCLASSINFOpdf
 
\else
  
\fi

\usepackage{amsthm}
\usepackage{makeidx}
\usepackage{examples}
\usepackage[pdftex]{graphicx}  
\usepackage{amssymb}
\usepackage{graphicx}
\usepackage{algorithm}
\usepackage{epstopdf}
\usepackage{color}
\usepackage{url}
\graphicspath{{../eps/}{../ps/}}
\usepackage{psfrag}    
\usepackage{algorithm}
\usepackage{algorithmic}
\usepackage{amsmath}

\IEEEoverridecommandlockouts
\begin{document}

\title{DNA Pen: A Tool for Drawing on a Molecular Canvas}

\author{
\IEEEauthorblockN{Arnav Goyal, Dixita Limbachiya, Shikhar Kumar Gupta, Foram Joshi, Sushant Pritmani, Akshita Sahai and \\
Manish K Gupta\\
Laboratory of Natural Information Processing,\\ 
Dhirubhai Ambani Institute of Information and Communication Technology\\
	Email: 200901040@daiict.ac.in, dlimbachiya@acm.org, gshikhri@gmail.com, foram.mj@gmail.com, \\
	sushant.pritmani@gmail.com, 200901063@daiict.ac.in, 
	m.k.gupta@ieee.org\\
}}

\maketitle

\begin{abstract}
DNA origami is an interdisciplinary area where DNA can be used as a building block for making useful stuff at nanoscale. This work presents an open source software DNA pen (based on the recent work of Peng Yin and his group) which can be used (using free hand and digital molecular canvas) to draw an object  at nanoscale. Software generates error free DNA sequences which can be used in the wet lab to create the object at the nanoscale. Using DNA pen we have drawn several objects including the map of India and sanskrit letter "Om" from free hand molecular canvas and digital letter DNA using digitized molecular canvas. 
\end{abstract}


\IEEEpeerreviewmaketitle
\section{Introduction}
In $1959$, in a famous lecture \cite{FL59} "There is a plenty of room at the bottom", Richard Feynman expressed a vision and desire to write at the nanoscale, which has inspired many people to attempt writing at nanoscale. In $1994$ in a seminal work  \cite{adleman94} Adleman pave the way for a new branch of DNA computing, by showing how to solve an instance of Hamiltonian path problem, using bunch of DNA strings (designed in a cleaver way) and the biotechnology lab methods. Subsequently Erik Winfree has shown that self-assembly of DNA is turing universal \cite{WIN2}. In $2006$, Paul Rothemund has shown how to fold the viral DNA into desired shape using helper DNA strands (known as staplers) giving birth to DNA origami \cite{rothemund2005design, Rothemund_2006, Castro_Kilchherr_Kim_Shiao_Wauer_Wortmann_Bathe_Dietz_2011,Nangreave2010608}. Many researchers have created interesting objects at nano-scale using DNA origami both at $2$D and $3$D \cite{Han15042011}. However scaling of these objects have been a major issue \cite{Nangreave2010608}.  Almost $50$ years from the date of Feynman's lecture, Pen Yin and his group has given a new direction to DNA origami by using DNA bricks made of arbitrary DNA strings \cite{wei2012complex, Ke30112012}. His team was successful in creating 2D and 3D objects using molecular canvas made up of DNA bricks. In this work, we present a software DNA pen that can be used to create 2D objects using DNA bricks of appropriate size. 

This paper is organized as follows. Section $2$ describes an outline of GUI and section $3$ provides a detailed description. Finally section $4$ provides the brief methods adopted for error free DNA sequences and section $5$ gives general remarks. Section $6$ provides a link for downloading the software and related materials.

\section{GRAPHICAL USER INTERFACE Overview}
The graphical user interface (GUI) for DNA Pen has been developed to enable the end user to get the desired DNA sequences by constructing different shapes on the molecular canvas.
There are three main functionalities of the software as described below:

\subsection{Free Hand Molecular Canvas}
\begin{figure}
 \includegraphics[scale=.25]{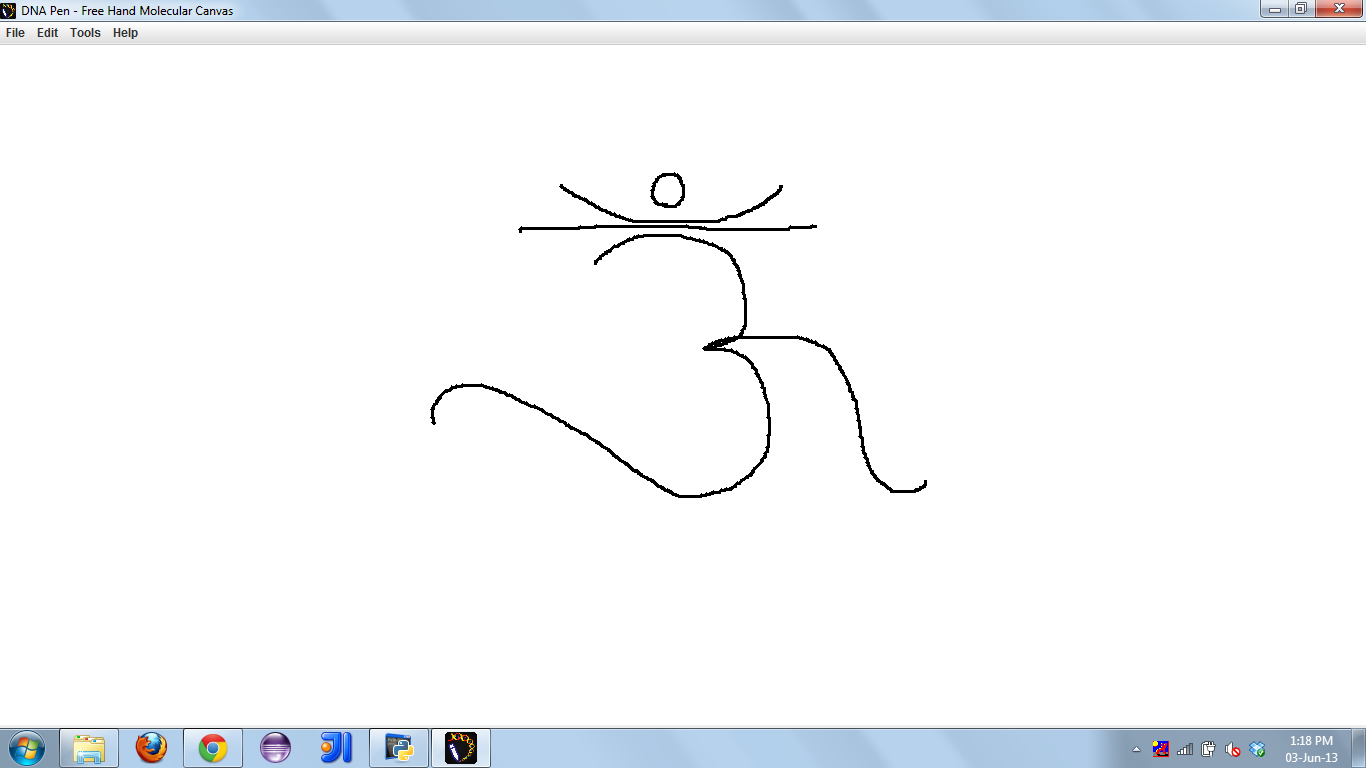}
\caption{OM drawn on Free Molecular Canvas}
\label{png1}
\end{figure}

It enables the user to draw any shape on the molecular canvas using the mouse as a paint brush. When the user starts drawing on this canvas, sequences of DNA bricks are added to the output file, which is then available to the user by calling the appropriate save function. This canvas allows the user to freely draw any shape user wants as shown in Figure \ref{png1}. These DNA sequences can be used to construct the nanoshapes in the lab.

\subsection{Digitized Molecular Canvas}
\begin{figure}
 \includegraphics[scale=.25]{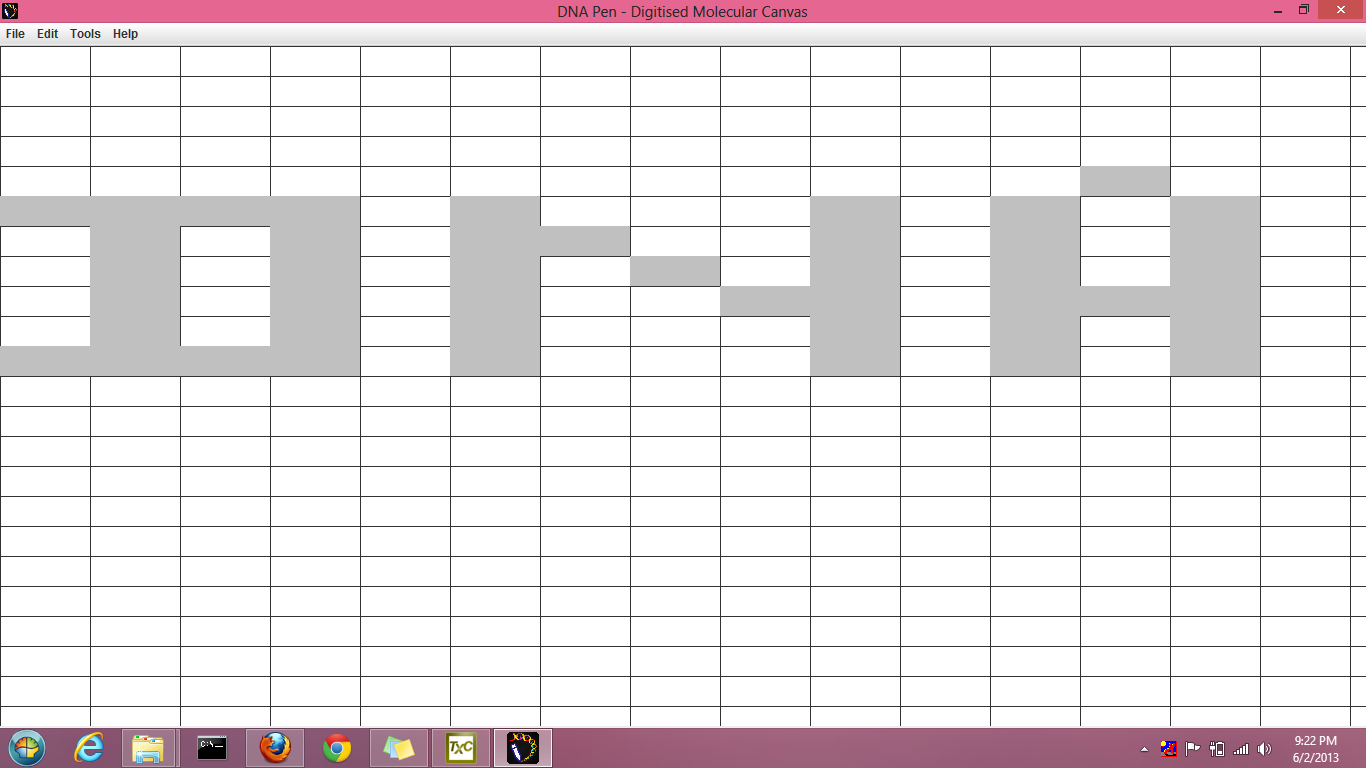}
\caption{{{DNA}} written on Digitized Molecular canvas}
\label{png2}
\end{figure}

All the functionalities in this canvas is similar to that of free hand molecular canvas except those mentioned below:

Each pixel in the Free Hand Molecular Canvas is a DNA brick in itself and thus the output contains the combinations of sequences of such bricks required to  construct the desired figure. Whereas, in the Digitized Molecular Canvas, each line, horizontal and vertical, represents a DNA sequence, which is a combination of the four domains. Thus the output contains combinations of sequences of these four domains rather than the sequences of an entire DNA brick. The Digitized Molecular Canvas allows the user to only draw along the horizontal and vertical lines unlike the free drawing property of the Free Hand Molecular canvas.
The user can draw only digitized objects as shown in Figure \ref{png2}. This canvas will give a more detailed output as instead of giving the entire DNA sequence of the DNA brick, it would give out specific combinations of the four domains. A user can deselect an already selected brick by simple clicking on it again.
To select the canvas user can go in Tools menu and select the desired canvas.

\subsection{DNA Brick Dimensions}
When user select the molecular canvas, the brick size $3 \times 7$ (nm) a  default dimension is generated. But if in case user wants to edit the dimension of the brick, software enable the user to do this by 'Edit Dimensions' option available in Edit menu. The DNA Pen software accepts two inputs from the user; brick height and brick width. The height and width are required to be multiples of 0.6 and 1.75 respectively \cite{wei2012complex}. If either of the height or width, or both do not satisfy the input conditions, they are automatically adjusted to meet the mathematical assumptions and a prompt is displayed to the user.  One can view some sample brick images  of different dimensions in the Tools menu. This feature of editing the brick dimension could be important for the scalibility of the nanoscale shape created.

\section{Detailed Description of GUI}

\subsection{Create New project}
 When user executes the software, the dialouge box with \textbf{Create new project} option will appear. User should specify the file name and browse the path to save the project at specific location. This will create the folder with the filename which will store all the output files. This option is also available in File menu.

\subsection{Save DNA Data}
When user creates the shape on molecular canvas, DNA sequences will be generated for the respective shape. This will include all the DNA sequences for each domain and the shape. One can save these sequences by clicking \textbf{SAVE DNA DATA} under the File menu. The output file with name DNAData$\_$filename will be generated. 

\subsection{Save Detailed DNA Data}
This option will create the DNA sequences with the coordinates for each shape created on the molecular canvas. Once the user has drawn the shape on the canvas and want to change the canvas, one can save this data and switch to other canvas.The file with name DetailedDNAData$\_$filename will be created.

\subsection{Save PDF}
\begin{figure*}%
 \includegraphics[scale=.8]{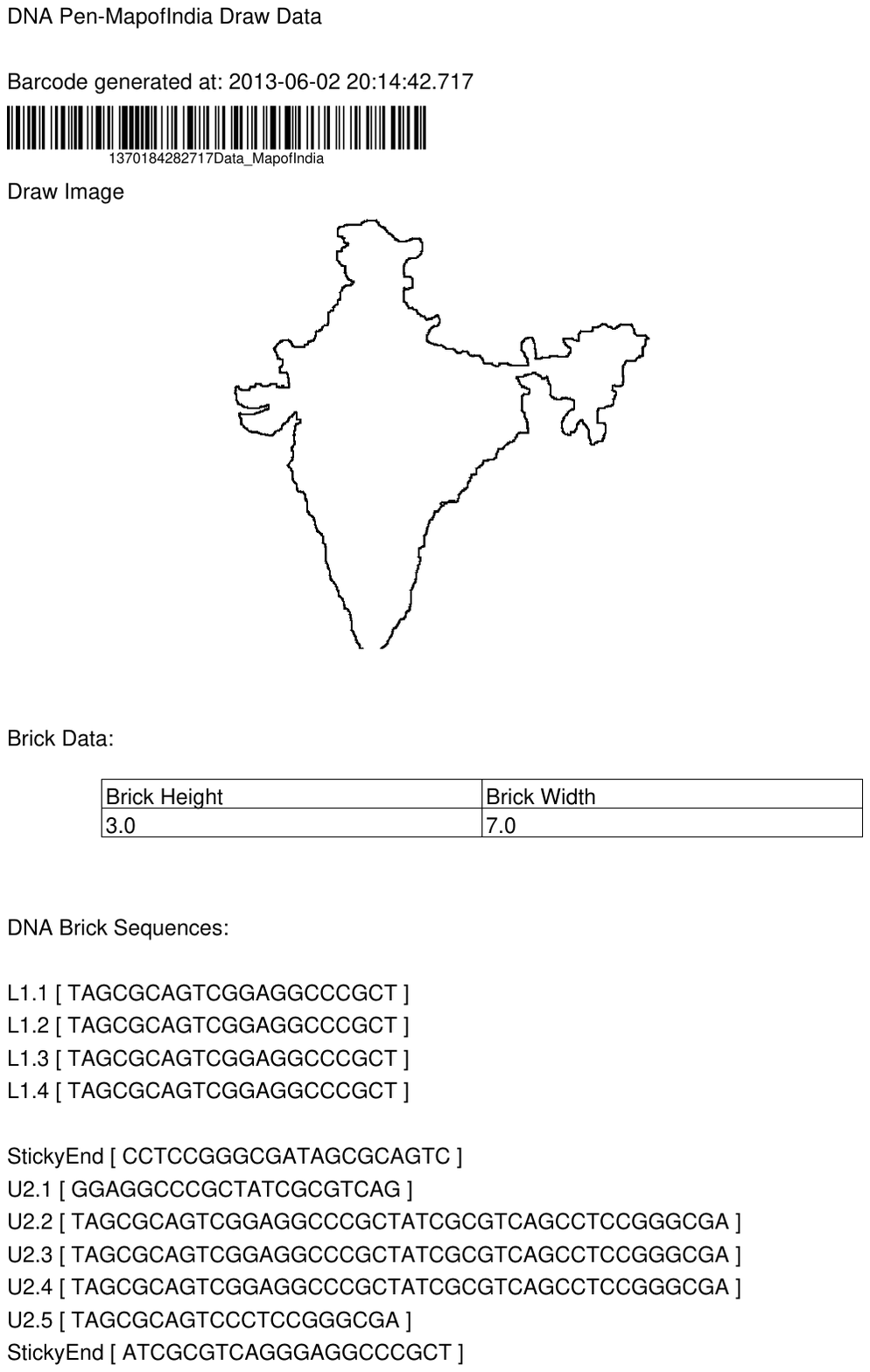}
 \caption{Output of a pdf file showing Map of India, brick dimensions and DNA Sequences arranged in bricks pattern}
\label{pdf}
\end{figure*}

This option includes all the output in one PDF. Once you draw the shape on the canvas, click on this option. It will include the barcode number which is unique to all the shapes drawn by the user.  This will help to uniquely identify the shapes drawn by him. It will include image of the structure designed on the canvas, dimension of the brick and DNA sequences for the shape. If the user has drawn the shape on the Free molecular canvas, the file FreeGridData$\_$filename will be generated. For the digitized canvas, file DigitizedDNAData$\_$filename will be generated. The file contains all the output in one page which will help user to understand it better. The map of India was created using DNApen on Free molecular canvas. (see Figure \ref{pdf}).

\subsection{Output}
\begin{figure*}
\includegraphics[scale=1.6]{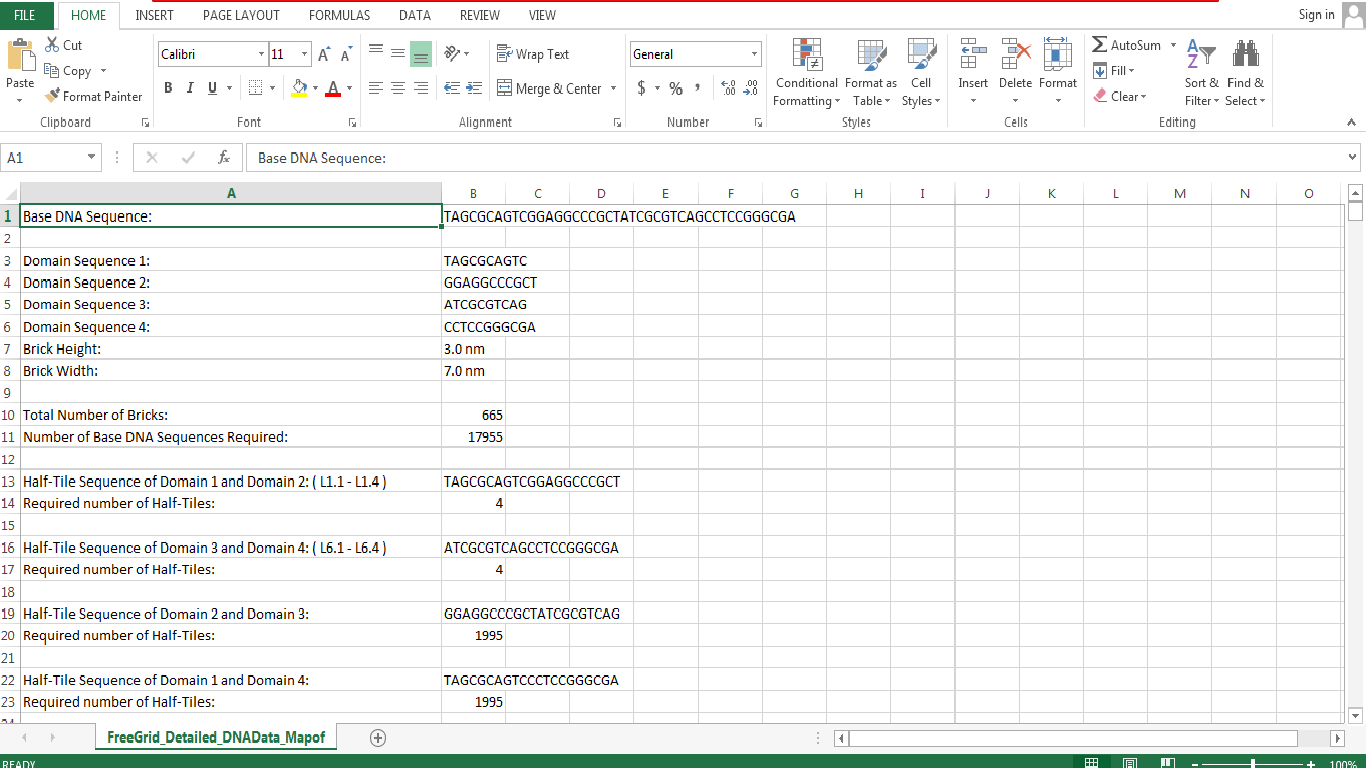}
\caption{Output DNA sequences for map of India and total nummber of DNA sequences required}
\label{xls}
\end{figure*}

Once you save the DNA data, the output generated is in the excel format. The project folder created will store all the output files.There will be three output as mentioned above in save section. DNAData$\_$filename gives the information about the base sequences required for the shape designed and includes the sequences for each domain in the DNA bricks, number of full and half tile for each domain and number of stick ends. DetailedDNAData$\_$filename  gives the coordinates for each base in the sequence on the basis of shape constructed on the molecular canvas. For digitized molecular canvas, the X co-ordinate denotes the west direction of the brick while the Y co-ordinate denotes the north direction of the brick. Figure \ref{xls} show the sample output file for the above shape created on the molecular canvas.  This output helps the researcher to design the shape with respect to arrangement of bricks DNA and the stick ends. 

\subsection{Window Change Event Handlers}
Whenever a user selects the \textbf{New} or the \textbf{Edit Brick Dimension} function and is currently drawing on either of the molecular canvases, the software prompts the user to save the draw data, if any, to be saved before switching to the new window. A similar prompt is shown when a user switches between the molecular canvases and there is some draw data to be saved. And also when a user wishes to create a new brick or edit the dimensions and had not yet saved the data of the already created brick. 

\subsection{Exit Event Handlers}
Many a times, users directly click on the \textbf{Red Cross} to exit the application or select the \textbf{Exit} option under the File menu. When such an event occurs, there might be a lot of unsaved data, which might be important to the end user.Thus, whenever the user tries to quit the application without saving the data, a prompt is shown which asks the user to save the unsaved draw data, if any, and the unsaved brick data, if any.

\subsection{Clearing the Molecular Canvas}
This will clear everything that user had drawn and start again from scratch. It will reset all the values to their original values.

\section{Error Correction Modules}

This module generates the error free DNA sequences. The generated sequences are checked against distance constriants \cite{DBLP:journals/jcb/MaratheCC01}. 
 If any sequence fails any of the tests, it is discarded and a new sequence is generated. This gives the stable, reliable and error free DNA sequences for nanoscale construction.

\section{Conclusion}
DNA Pen has been developed for the sole-purpose of generating a user-friendly, interactive environment for users to envisage their DNA structures, and get the actual DNA sequences required to make the physical strucutures. With the generation of error free DNA sequences by error correction module, the resultant sequences are stable and reliable.Thus the output sequences can be experimentally used to make the nanoscale architectures with specified brick design. In the future, we expect to enhance the functionalities of the software and enable the user to draw 3-dimensional objects on the molecular canvas. 
\section{SOFTWARE AVAILABILITY}
 The software (source code), installers for Mac and Windows, user manual, product demo and other related materials  can be downloaded from http://www.guptalab.org/dnapen/.
\section{ACKNOWLEDGEMENT}
The authors would also like to thank Priyanka Shukla and Sonam Jain for providing the module to generate error free DNA sequences and Bruno Lowagie (http://itextpdf.com) whose free open source libraries have been used for generating bar codes and pdf. 


\bibliography{dnapenref}
\bibliographystyle{IEEEtran} 

\end{document}